\documentclass[twocolumn,showpacs,showkeys,preprintnumbers,superscriptaddress,amsmath,floatfix,amssymb,secnumarabic,nofootinbib]{revtex4}
\usepackage[colorlinks=true]{hyperref}
\usepackage{graphicx}
\usepackage{braket}
\usepackage{amsmath}
\usepackage{float}
\usepackage{caption}
\usepackage{subcaption}
\usepackage{color}

\def\({\left(}
\def\){\right)}
\def\[{\left[}
\def\]{\right]}

\newcommand{\ev}{\, {\rm eV}}

\newcommand{\lr}[1]{ \left( #1 \right) }
\newcommand{\lrs}[1]{ \left[ #1 \right] }

\newcommand{\tr}{ {\rm Tr} \, }

\newcommand{\beq} {\begin{eqnarray}}
\newcommand{\eeq} {\end{eqnarray}}

\newcommand{\comment}[1]{}

\begin{document}
\sloppy

\title{Many-body effects on graphene conductivity: Quantum Monte Carlo calculations}

\author{D.~L.~Boyda}
\email{boyda_d@mail.ru}
\affiliation{Far Eastern Federal University, Vladivostok, 690091 Russia}
\affiliation{Institute of Theoretical and Experimental Physics, 117259
Moscow, Russia}

\author{V.~V.~Braguta}
\email{victor.v.braguta@gmail.com}
\affiliation{Institute of Theoretical and Experimental Physics, 117259
Moscow, Russia}
\affiliation{School of Biomedicine, Far Eastern Federal
University, Sukhanova 8, Vladivostok, 690950 Russia}
\affiliation{Moscow Institute of Physics and Technology, Institutskii
per. 9, Dolgoprudny, Moscow Region, 141700 Russia}

\author{M.~I.~Katsnelson}
\email{m.katsnelson@science.ru.nl}
\affiliation{Radboud University, Institute for Molecules and Materials,
Heyendaalseweg 135, NL-6525AJ Nijmegen, The Netherlands}
\affiliation{Ural Federal University, Theoretical Physics and Applied Mathematics Department, Mira Str. 19,  620002 Ekaterinburg, Russia}

\author{M.~V.~Ulybyshev}
\email{ Maksim.Ulybyshev@physik.uni-regensburg.de}
\affiliation{Institute of Theoretical Physics, University of Regensburg,
D-93053 Germany, Regensburg, Universitatsstrasse 31}
\affiliation{Institute for Theoretical Problems of Microphysics, Moscow State University, Moscow, 119899 Russia}

\begin{abstract}
Optical conductivity of graphene is studied using Quantum Monte Carlo calculations. We start from Euclidean current-current correlator and extract $\sigma (\omega)$ from Green-Kubo relations using Backus-Gilbert method.  Calculations were performed  both for long-range interactions and taking into account only contact term. In both cases we vary interaction strength and study its influence on optical conductivity. We compare our results with previous theoretical calculations choosing $\omega \approx \kappa$ thus working in the region of the plateau in $\sigma(\omega)$ which corresponds to optical conductivity of Dirac quasiparticles.
No dependence of optical conductivity on interaction strength is observed unless we approach antiferromagnetic phase transition in case of artificially enhanced contact term. Our results strongly support previous  theoretical  studies claimed very weak regularization of graphene conductivity.

\end{abstract}

\maketitle

\section{Introduction}

Low-energy electronic properties of graphene in single-particle approximation are determined by fermionic excitations in $p_z$ band with
a linear relativistic dispersion relation $\mathcal{E} = v_F |p|$ and the Fermi
velocity $v_F \simeq c/300$ \cite{wallace,mcclure,semenoff,Novoselov:05:1,zhang}. Due to the smallness of the Fermi velocity $v_F \ll c$
magnetic interaction between fermion excitations and retardation effects are negligible.
As a result, the interaction between quasiparticles is instantaneous Coulomb law with
effective coupling constant
\beq
\alpha = \frac 1 {\epsilon} \frac {c} {v_F}  \frac {1} {137}
\eeq
where $\epsilon$ is dielectric permittivity of surrounding media. For suspended
graphene the effective coupling constant is sufficiently large $\alpha_{0} \approx 2$.
Quite strong interaction between fermion excitations in graphene results in a rich variety of phenomena \cite{Kotov:2}. In particular, ``relativistic'' invariance of the theory is destroyed by renormalization of the Fermi velocity which turns out to be strongly $k$-dependent, $k$ is the wave vector. This prediction \cite{Gonzalez:1993uz} has been recently confirmed experimentally \cite{Elias,PNAS}. Accurate procedure of mapping of full electron Hamiltonian onto $p_z$ subspace results in a noticeable screening of effective interaction potential at intermediate distances in comparison with the bare Coulomb \cite{Wehling} which does not change qualitatively perturbative results \cite{Braguta2015} but shifts the point of semimetal-insulator transition making freely suspended graphene semimetallic \cite{Ulybyshev:2013swa}.

An interesting observable for which one can expect sizable renormalization is the optical conductivity $\sigma(\omega)$.
The $\sigma(\omega)$ describes electric current in graphene resulting from external electric field $j(\omega)=\sigma(\omega) E(\omega)$.
In the non-interacting theory of Dirac quasiparticles the conductivity does not depend on frequency in the limit of zero temperature and zero doping. It equals to\cite{Katsnelson}  $\sigma_0 = \frac {1} {4}$ in the units of $\frac {e^2}{\hbar}$.
The difference of the $\sigma(\omega)$ from the value $\sigma_0$ can be attributed to
the interaction between quasiparticles. Experimental data on the optical
conductivity did not find \cite{Li, Mak, Nair} any deviation from the result of single-particle
theory, within experimental error bar, a few percent. This seems to be surprising keeping in mind that many-body effects change dramatically the single-electron spectrum and electron compressibility \cite{Elias,PNAS}. 

There are a lot of papers devoted to the leading-order perturbative correction to the value $\sigma_0$ \cite{Mishchenko:1,Kotikov, Herbut, Sheehy, Juricic, Mastropietro, Katsnelson2008, Rosenstein, Link2015}
which present different results. In general, all these papers can be divided into two groups. First of all, it was proven in Ref. \cite{Mastropietro} that non-renormalization of conductivity in the limit of zero frequencies for weak on-site interaction is a rigorous result following from symmetry required exact cancellation of self-energy and vertex contributions. This work was based on renormalization group treatment of Hubbard model on honeycomb lattice. Thus, long-range interactions were not taken into account. Earlier, the absence of renormalization was demonstrated phenomenologically based on Fermi-liquid theory \cite{Katsnelson2008}. Whereas the system of electrons with contact interaction at the honeycomb lattice is probably Fermi liguid, for the case of long-range interactions this is not the case \cite{Kotov:2}. In the second group of papers the influence of long-range Coulomb tail on optical conductivity of massless Dirac fermions was studied. The renormalization of optical conductivity was found:
\beq
\frac {\sigma(\omega)} {\sigma_0} = 1+ C \alpha_{eff} +O (\alpha_{eff}^2),
\label{sigma_renorm}
\eeq
with $\alpha_{eff}$ logarithmically dependent on frequency, but  the value of the constant $C$ was disputable. Since almost all of these papers were based on Dirac cones approximation,  $C$ was essentially dependent on the regularization scheme used in calculations.
The authors of Refs. \cite{Herbut, Juricic} claimed rather large renormalization with $C=0.2...0.5$. On the other hand, Mishchenko \cite{Mishchenko:1} and Kotikov \cite{Kotikov} observed very small constant $C \approx 0.01$. The authors of Ref. \cite{Rosenstein} obtained $C\approx0.26$ working on honeycomb lattice thus using a natural cutoff.  In the most recent paper \cite{Link2015} the authors calculated  $C$ taking into account non-zero size of atomic orbitals  in interaction term. Similar to \cite{Mishchenko:1} they obtained  very small renormalization.

This short review shows that renormalization of conductivity in graphene is still an open, and quite controversial, theoretical question. It seems that experiment supports small corrections to conductivity \cite{Li, Mak, Nair}, thus there should exist  a considerable cancellation between self-energy and vertex corrections. From the other side, there are not any clear physical reasons which stay behind this
cancellation. In particular, this is not the case for the other response function, electron compressibility, where self-energy corrections are dominant \cite{Kotov:2,PNAS}. What is more important it is not clear if this cancellation takes place
for higher order perturbation theory.  So, the absence or considerable suppression of the many-body effects in optical conductivity is important theoretical puzzle.

In this paper we are going to address the question of many-body effects in the optical
conductivity using Quantum Monte-Carlo simulation which fully accounts interactions
between quasiparticles in a nonperturbative way. In the second section we describe our model and give a short review of Quantum Monte Carlo algorithm. Since the calculation of optical conductivity is based on linear response theory, we describe also the calculation of current-current correlator.  The third section is devoted to the analytical continuation methods employed to obtain real-time conductivity  $\sigma(\omega)$. We present the final dependence of conductivity on interaction strength both for model with long-range and short-range interaction and give a short discussion of the results in conclusion.

\section{Model description and calculation of current-current correlator}

Our model Hamiltonian consists of tight-binding term and interaction part which describes the full electrostatic interaction between quasiparticles :
\begin{eqnarray}
\label{tbHam}
 \hat{H}= - \kappa \sum\limits_{<x,y>} \lr{ \hat{a}^{\dag}_{y, \uparrow} \hat{a}_{x, \uparrow}
+  \hat{a}^{\dag}_{y, \downarrow} \hat{a}_{x, \downarrow} + h.c.}
 \nonumber \\ +
 \sum\limits_{x=\{1,\xi\}} m (\hat{a}^{\dag}_{x, \uparrow} \hat{a}_{x, \uparrow} - \hat{a}^{\dag}_{x, \downarrow} \hat{a}_{x, \downarrow} ) \nonumber \\ - \sum\limits_{x=\{2,\xi\}} m (\hat{a}^{\dag}_{x, \uparrow} \hat{a}_{x, \uparrow} - \hat{a}^{\dag}_{x, \downarrow} \hat{a}_{x, \downarrow} ) \nonumber \\ +{1 \over 2} \, \sum\limits_{x,y} V_{xy} \hat{q}_x \hat{q}_y,
\end{eqnarray}
where $\kappa = 2.7 \ev$ is hopping between nearest-neighbours, $\hat{a}^{\dag}_{x, \uparrow}$, $\hat{a}_{x, \uparrow}$ and $\hat{a}^{\dag}_{x, \downarrow}$, $\hat{a}_{x, \downarrow}$ are creation/annihilation operators for spin up and spin down electrons at $\pi$-orbitals. Spatial index $x=\{s, \xi\}$ consists of sublattice index $s=1,2$ and two-dimensional coordinate $\xi=\{\xi_1, \xi_2\}$ of the unit cell in rhombic lattice. Periodical boundary conditions are imposed in both spatial directions. The mass term has different sign at different sublattices. According to our algorithm, we should add it to eliminate zero modes from fermionic determinant. Of course it means that we should check the dependence of final results on the value of mass and this check is indeed performed in subsequent calculations.

The matrix $V_{xy}$ is electrostatic interaction potential between sites with coordinates $x$ and $y$ and $\hat{q}_x = \hat{a}^{\dag}_{x, \uparrow} \hat{a}_{x, \uparrow} + \hat{a}^{\dag}_{x, \downarrow} \hat{a}_{x, \downarrow} - 1$ is the operator of electric charge at lattice site $x$. We use two sets of interaction potentials. The first  one (``long range interaction'')  consists of phenomenological potentials calculated by the constrained RPA method \cite{Wehling} at small distances. It continues at large distances by ordinary Coulomb  $V_{xy} \sim 1 / r_{xy}$. In general this set up corresponds to suspended graphene, further details can be found in Ref. \cite{Ulybyshev:2013swa}.  In the second set up (``short range interaction'')  we keep only on-site interaction equal to 9.3 eV. This number, again, accords to Ref. \cite{Wehling}. In order to study dependence on interaction strength in both cases we used rescaling of potentials where all of them were uniformly divided by dielectric permittivity of surrounding media $\epsilon$. Thus suspended graphene corresponds to the point $\epsilon=1$ on the plots for long range interaction.

\begin{figure}[t!]
        \centering
        \includegraphics[scale=0.27, angle=270]{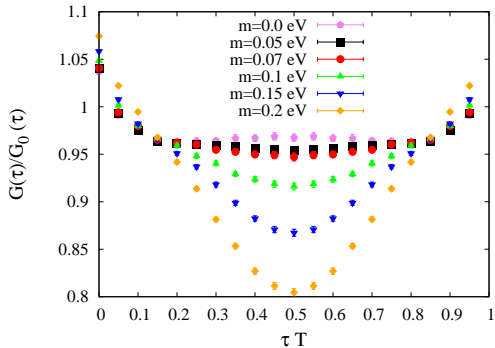}
        \caption{The ratio of current-current correlator calculated in presence of interaction and for free fermions ($G(\tau)/G_0(\tau)$) as a function of Euclidean time $\tau$ for different bare masses.  Temperature is equal to $T=0.5$ eV  and  lattice size is  $24^2\times20$. Interaction strength corresponds to suspended graphene.}
        \label{fig:bare_corr}
\end{figure}

\begin{figure}[t!]
\centering
\begin{subfigure}{.5\textwidth}
  \centering
  \includegraphics[scale=0.27, angle=270]{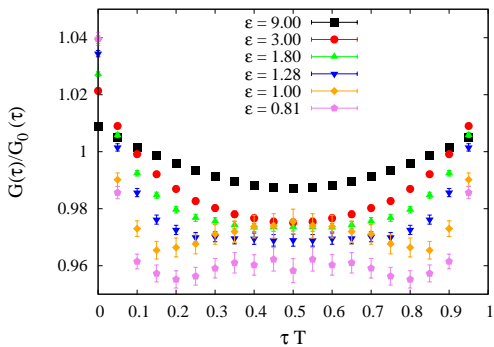}
  \caption{The plot corresponds to the model with long-range interaction.}
  \label{fig:correlators1}
\end{subfigure}
\begin{subfigure}{.5\textwidth}
  \centering
  \includegraphics[scale=0.27, angle=270]{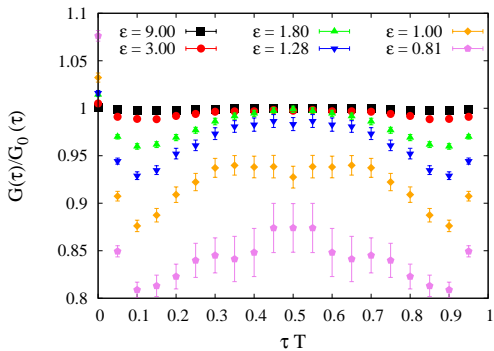}
  \caption{The plot corresponds to the model with pure on-site interaction.}
  \label{fig:correlators2}
\end{subfigure}
\caption{The ratio of current-current correlator calculated in presence of interaction  ($G(\tau)$) and the one for free fermions ($G_0(\tau)$) as a function of Euclidean time $\tau$ for different interaction strengths.  Temperature is equal to $T=0.5$ eV  and  lattice size is  $24^2\times20$. Both plots show the data for zero bare mass.}
\label{fig:correlators}
\end{figure}

All calculations were performed using Hybrid Monte-Carlo algorithm. Details of the algorithm are described in the papers \cite{Buividovich:2012nx, Ulybyshev:2013swa, Lorenz2014}. 
The method is based on Suzuki-Trotter decomposition. Partition function $\exp (-\beta \hat H)$ is represented  in the form of a functional integral in Euclidean time. Inverse temperature is equal to number of time slices multiplied by the step in Euclidean time: $\delta \tau N_t=\beta=1/T$. Since the algorithm needs for fermionic fields to be integrated out, we eliminate all four-fermionic terms in full Hamiltonian  using Hubbard-Stratonovich transformation. The final form of the partition function can be written as:
\begin{eqnarray}
\label{PartFunc2}
 \tr e^{-\beta \hat{H}} \cong \int \mathcal{D}\varphi_{x,n} e^{-S\lrs{\varphi_{x,n}}}
 |\det{M\lrs{\varphi_{x,n}}}|^2.
\end{eqnarray}
$\varphi_{x,n}$ is the Hubbard-Stratonovich field for timeslice $n$ and spatial coordinate $x$. Particular form of fermionic operator $M$ is described in Ref. \cite{Ulybyshev:2013swa}. The absence of sign problem (appearance of the squared modulus of the determinant) is guaranteed by the particle-hole symmetry in graphene at neutrality point. Action for hubbard field $S\lrs{\varphi_{x,n}}$ is also positively defined quadratic form for all variants of electron-electron interaction used in our paper. Thus we can generate configurations of $\varphi_{x,n}$ by the Monte-Carlo method using the weight (\ref{PartFunc2}) and calculate physical quantities as averages over  generated configurations.

 We extract conductivity from Euclidean correlation function of electromagnetic currents:
\beq
G(\tau) = \frac {T_{bc}} {3 \sqrt 3 N_s } \sum_{\xi}  \frac {\mbox{Tr}  \left( { e^{-\beta \hat H} \hat J_{b}(\xi) e^{-\tau \hat H} \hat J_{c}(\xi) e^{\tau \hat H} }\right)  }{\mbox{Tr} \, e^{-\beta \hat H}} ,
\label{correlation_f}
\eeq
where $N_s$ is the number of unit cells in our sample, summation over repeated indexes is implied. $\hat J_{b}(\xi)$ is the operator of electromagnetic current flowing  from the site in the first sublattice with coordinate $x=\{1,\xi\}$  towards one of its nearest neighbours. There are three nearest neighbours and 3 possible directions of the current: $b=1,2,3$. Matrix $T_{bc}$ is defined as follows:
\beq
T_{bc}=d^2 \left\{ {  {1, \quad b=c}  \atop  {-1/2, \quad b \neq c}  }  \right.  ,
\label{matrix_T}
\eeq
where $d=0.124\, \mbox{nm}$ is the distance between nearest neighbours in graphene lattice.
We need in calculation the explicit form of electromagnetic current operator:
\beq
\hat J_b (\xi) =  \sum_{\sigma=\uparrow, \downarrow} \hat J_{\sigma, b} (\xi) = i \kappa \sum_{\sigma=\uparrow, \downarrow}  \hat a^{\dag}_{\sigma, x} \hat a_{\sigma,y} + h.c., 
\eeq
where $x=\{1, \xi \}$, $y=\{2, \xi+\rho_b \}$ and shifts in rhombic lattice are defined as follows: $\rho_1=\{0,0\}$, $\rho_2=\{-1,1\}$, $\rho_3=\{-1,0\}$. 
The final form of the current-current correlator used in Monte Caro procedure can be written in terms of some combinations of the lattice fermionic propagator $\Gamma(x,\tau, x,\tau') = M^{-1} (x,\tau, x,\tau')$ (see \cite{Buividovich:2012nx} for derivation):
\begin{widetext}
\beq
G(\tau) = - \frac {2 T_{bc}} {3 \sqrt 3 N_s} \langle \mbox{Re}  \sum_{x_1,x_2,y_1,y_2}\bigl [ j_b(y_1,x_1) \Gamma(x_1,0, x_2,\tau) j_c(x_2, y_2) \Gamma (y_2,\tau,y_1, 0) \bigr ] \rangle +
\nonumber \\
   \frac {4 T_{bc}} {3 \sqrt 3 N_s} \langle \mbox{Re} \; \sum_{x,y} \bigl [ j_b(y, x) \Gamma(x,0, y, 0) \bigr ]
\cdot \mbox{Re}  \sum_{x,y} \bigl [ j_c(y,x) \Gamma(x,\tau, y, \tau) \bigr ]
 \rangle.
\label{lattice_correlation}
\eeq
\end{widetext}
Here $j_b$ is current vertex:
\beq
\sum_{\xi} \hat J_{\sigma, b} (\xi) = \sum_{ x=\{s, \xi \} \atop y=\{s', \xi' \}}  \hat a^{\dag}_{\sigma, x} \hat a_{\sigma, y}  j_b (x,y),
\label{j_matrix}
\eeq
and $\langle ... \rangle$ is average over configurations of hubbard field generated according to the statistical weight (\ref{PartFunc2}).

The first term in  expression (\ref{lattice_correlation}) is named as connected part and
the second term is named as disconnected part of the correlation function $G(\tau)$.  Connected part was calculated using standard stochastic estimator approach \cite{degrand}.  
The disconnected part is very noisy thus leading to very large uncertainty in
the Monte-Carlo simulation. Fortunately one can argue that it is small as compared to the
connected part.  For $\tau \neq 0$ the disconnected part
is the average of the operators which are located at different Euclidean
time slices (see equation (\ref{lattice_correlation})). To get nonzero result
for the disconnected contribution the operators at different $\tau$--slices
must interact. However, the interaction between different $\tau$--slices is suppressed
by one power of the Fermi velocity. Thus the contribution of the disconnected diagrams
 is very small in comparison with connected ones and we can safely neglect them in the correlation function (\ref{lattice_correlation}). Explicit check of this fact was done in the paper \cite{Buividovich:2012nx}. In particular, we refer to the the figure 11. Profile of electron-electron interaction in our paper differs from the one used in \cite{Buividovich:2012nx}, so we can not compare results with the same $\varepsilon$ directly. Nevertheless, it's possible to use the antiferromagnetic phase transition as a reference point. Since our setup of electron-electron interaction corresponds to the points in the semi-metallic phase away from the phase transition, we can use data sets also for semi-metallic phase ($\varepsilon=10$ and $\varepsilon=4$) at the figure 11 in \cite{Buividovich:2012nx} as an illustration of the fact that disconnected part of current-current correlator is much less than the connected one. Actually, it's even impossible to distinguish disconnected part from zero because of statistical errors.  

Technically, we carry out the measurements of $G(\tau)$ at lattice with $N_s=24^2$ cells and 20 Euclidean timeslices ($\delta \tau = 0.1 \mbox{eV}^{-1}$), which corresponds to the temperature $T=0.5$ eV. The data are obtained for the set of 5 fermion masses
$m=0.05$, 0.07, 0.1, 0.15, 0.2 eV. 
In order to study systematic errors appeared due to the finite volume and finite discretization step in Euclidean time direction
 we also measured the correlation function $G(\tau)$
 at lattice with $N_s=18^2$, $N_t=20$  and lattice with  $N_s=24^2$, $N_t=40$ ($\delta \tau$ is two times smaller in the latter case). The results of the
measurements show that the finite volume and finite discretization errors  are of order of
statistical uncertainty $\sim 0.3 \%$.

\begin{figure}
\centering
\begin{subfigure}{.5\textwidth}
  \centering
  \hspace*{-0.9cm}\includegraphics[scale=0.29, angle=0, trim={0cm 0 0 0 0}, clip]{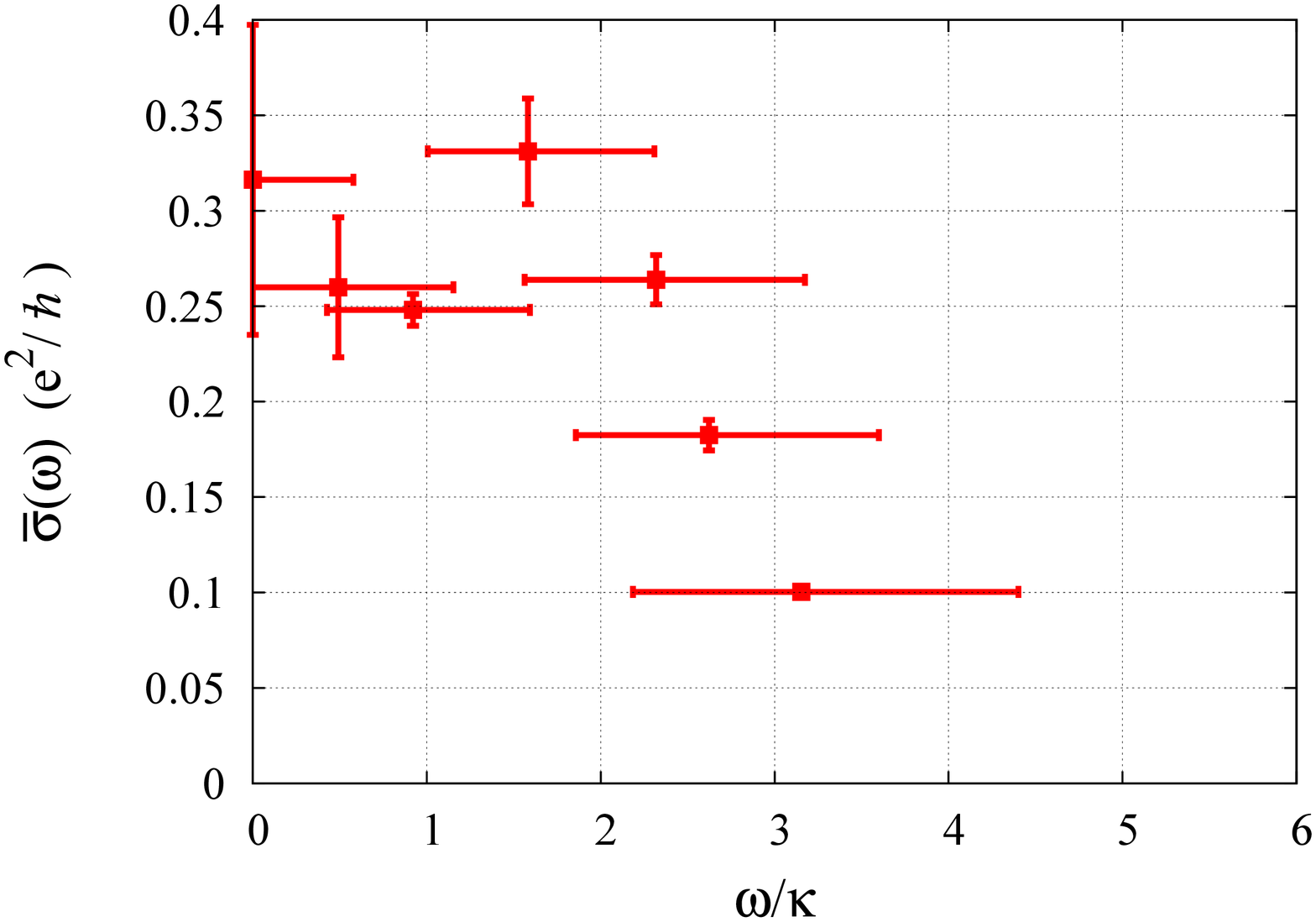}
  \caption{$\bar \sigma(\omega)$ extracted from Green-Kubo relations. Vertical error bar is statistical uncertainty. Horizontal error bar shows the width of corresponding resolution function. It is calculated as a width of the peak at half of its maximum height. Central point is placed at $\omega_c$: maximum of corresponding resolution function. }
  \label{fig:cond_free}
\end{subfigure}
\begin{subfigure}{.5\textwidth}
  \centering
  \includegraphics[scale=0.27, angle=270]{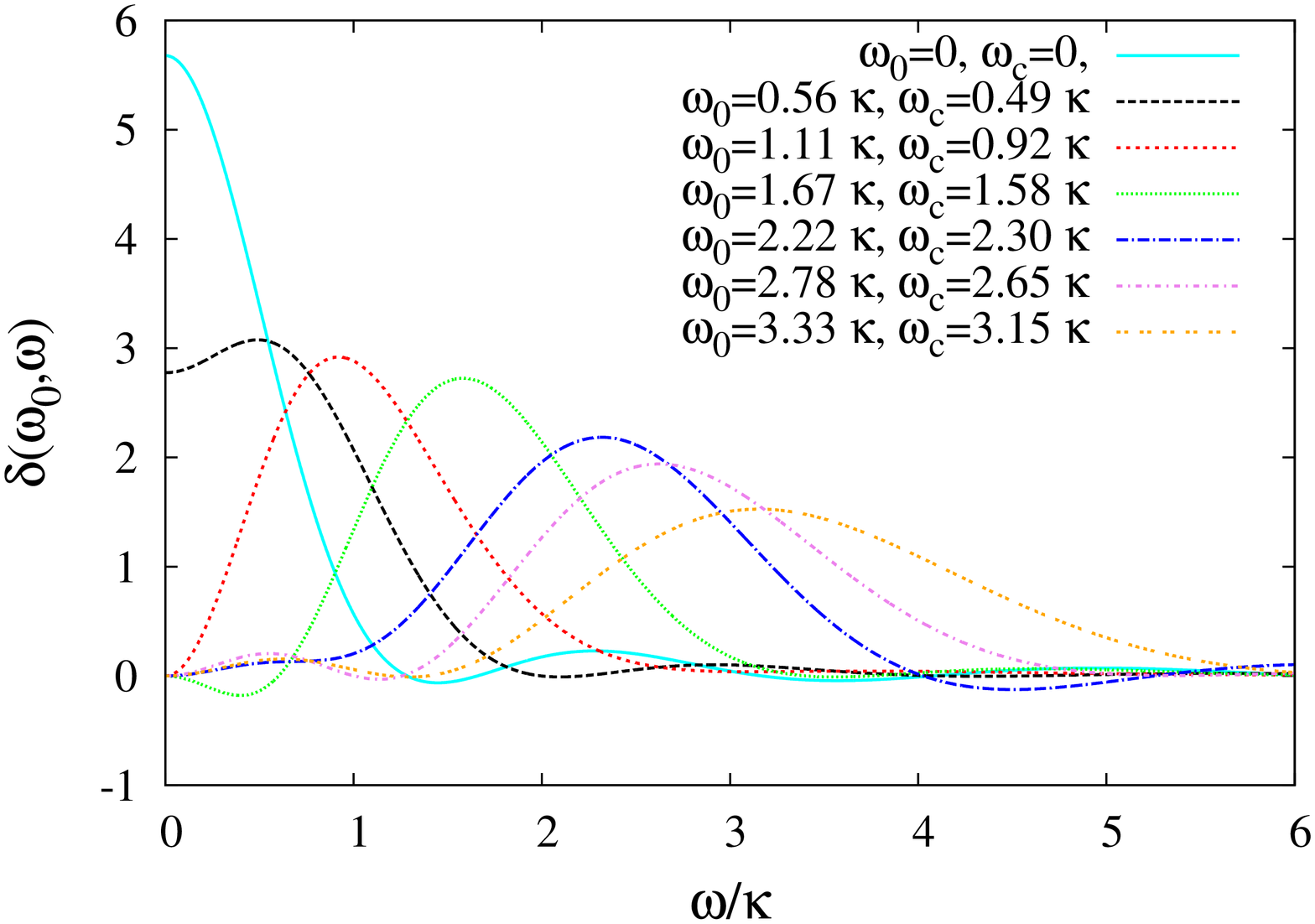}
  \caption{Resolution functions $\delta(\omega_0, \omega)$.}
  \label{fig:delta_free}
\end{subfigure}
\caption{ Results of test analytical continuation for free fermions: $\bar \sigma(\omega)$ and corresponding resolution functions. Regularization is absent: $\lambda=1$. Bare mass term is introduced:  $m=0.05$ eV. $\kappa=2.7$ eV is hopping between nearest neighbours.}
 \label{fig:conductivity_omega_free}
\end{figure}

\begin{figure}
\centering
\begin{subfigure}{.5\textwidth}
  \centering
  \hspace*{-0.9cm}\includegraphics[scale=0.29, angle=0]{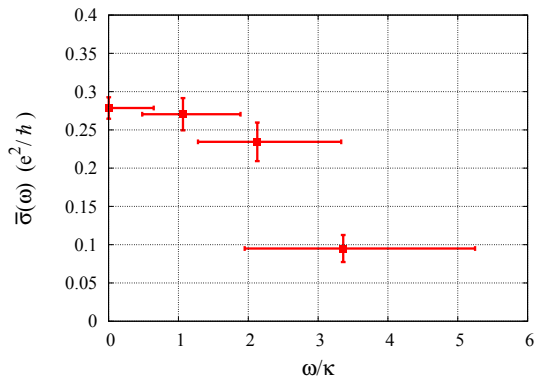}
  \caption{$\bar \sigma(\omega)$ extracted from Green-Kubo relations.  Vertical error bar is statistical uncertainty. Horizontal error bar shows the width of corresponding resolution function. It is calculated as a width of the peak at half of its maximum height. Central point is placed at $\omega_c$: maximum of corresponding resolution function.}
  \label{fig:cond_int}
\end{subfigure}
\begin{subfigure}{.5\textwidth}
  \centering
  \includegraphics[scale=0.27, angle=270]{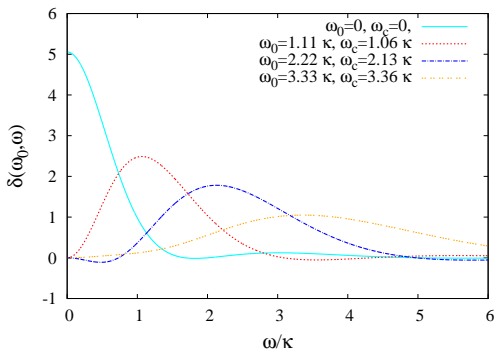}
  \caption{Resolution functions $\delta(\omega_0, \omega)$.}
  \label{fig:delta_int}
\end{subfigure}
\caption{ Real calculation for suspended graphene (long-range interaction with $\epsilon=1$). In this case regularization is imposed: $\lambda=1-4\times 10^{-5}$. Bare mass term is introduced:  $m=0.05$ eV. $\kappa=2.7$ eV is hopping between nearest neighbours.}
 \label{fig:conductivity_omega_int}
\end{figure}

The bare data for current-current correlator in suspended graphene is presented in fig. \ref{fig:bare_corr}. This is the  ratio $G(\tau)/G_0(\tau)$, where $G_0(\tau)$ is the correlator calculated for free fermions in the massless limit.  Data for zero bare mass is obtained via extrapolation by the function $G_{m}(\tau)=A(\tau)+B(\tau) \cdot m^2 + C(\tau) \cdot m^4$. 
It should be noted that renormalization of the $G(\tau)$ strongly depends on the fermion bare mass $m$. The smaller the fermion mass used in the calculation the smaller the change of the correlation function $G(\tau)$ due to the interaction.

 The ratio $G(\tau)/G_0(\tau)$ for different values of interaction is shown in fig. \ref{fig:correlators} for the massless limit. 
One can see that renormalization of the $G(\tau)$ is not very large. It doesn't exceed $5\%$ for long range interaction. It becomes noticeable only for the largest values of  short-range interaction which are close to antiferromagnetic phase transition (see below).

\section{Analytical continuation and calculation of conductivity}

The next stage is transformation of Euclidean current-current correlator to real-frequency optical conductivity via Green-Kubo relations:
\beq
 G(\tau) = \int_0^\infty  \sigma(\omega) K(\tau, \omega) d\omega, \\
 K(\tau, \omega)=\frac {\omega \cosh (\omega(\beta/2 -\tau))} {\pi \sinh(\omega \beta/2)}.
\eeq
We used  Backus-Gilbert method, recently adopted for similar tasks in lattice quantum chromodynamics \cite{Meyer2015}. 
According to this method we obtain convolution of optical conductivity with resolution functions $\delta(\omega_0, \omega)$:
\beq
\bar \sigma(\omega_0) = \int_0^\infty  \delta(\omega_0, \omega) \sigma(\omega) d\omega.
\eeq
Resolution functions are defined as linear combinations:
\beq
\delta(\omega_0, \omega) = \sum_{j=1}^{N_t/2} q_j(\omega_0) K(j \delta\tau, \omega),
\label{delta_def}
\eeq
where coefficients $q_j(\omega_0)$ can be obtained from the minimization of the  width of the resolution function. The width is defined as $D=\int_0^\infty  (\omega-\omega_0)^2 \delta(\omega_0, \omega) d\omega$. This relation guarantees the minimum width of resolution function with center close (but not exactly equal to) $\omega_0$. We also take into account the normalization condition $ \int_0^\infty  \delta(\omega_0, \omega) =1$. The final formula for conductivity and coefficients in (\ref{delta_def}) has the following form (see \cite{Meyer2015} for details):  
  \beq
 \bar \sigma(\omega_0)=\sum_{j=1}^{N_t/2} q_j(\omega_0) G(j \delta\tau), \\
  q_j(\omega_0)=\frac{W(\omega_0)^{-1}_{jk} R_k} {R_n W(\omega_0)^{-1}_{nm} R_m},
  \eeq
  where $W(\omega_0)_{jk}=\lambda \int_0^\infty (\omega-\omega_0)^2 K(j \delta \tau, \omega) K (k \delta \tau, \omega) d \omega + (1-\lambda) S_{jk}$ and $R_n=\int_0^\infty K( n\delta \tau, \omega) d \omega$. $S_{jk}$ is covariance matrix of current-current correlator. Summation over repeated indices is implied.
Parameter $\lambda$ is used to regularize poorly-defined matrix $W_{jk}$. It rules the width of resolution functions. If  quality of data is good enough, we can work without any regularization ($\lambda=1$) and obtain minimal width of the functions $\delta(\omega_0, \omega)$.  With decreasing of $\lambda$ the width of resolution functions increases but algorithm becomes more tolerant to statistical errors in $G(\tau)$. The examples of resolution functions are plotted in the figure \ref{fig:delta_free} for $\lambda=1$ (without regularization) and in the figure \ref{fig:delta_int} for $1-\lambda=1\times10^{-5}$. Value of $\omega_c$ gives the coordinate of maximum of corresponding resolution function. One can see that it doesn't coincide with $\omega_0$ precisely, but is very close to it. $\omega_c$ is used as x-coordinate in the plots for $\sigma(\omega)$.

Results of test calculation of conductivity for free fermions are shown in the figure  \ref{fig:cond_free}. Even in calculation of current-current correlator for free fermions we used the same numerical procedure (stochastic estimator) \cite{degrand} as for real calculation in interacting model. Thus we have some residual statistical uncertainties which give us a possibility to check stability of the method against errors in input data. It appears that test calculation for free fermions can be performed even without any regularization, so $\lambda=1$  for all data in the figure \ref{fig:conductivity_omega_free}.

\begin{figure}[t!]
        \centering
        \includegraphics[scale=0.27, angle=0]{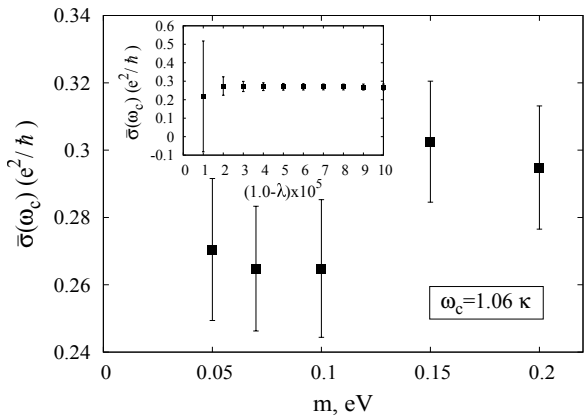}
        \caption{ Dependence of  $\bar \sigma (\omega_c)|_{\omega_c=1.06 \kappa}$ on bare mass for $\lambda=1-4\times 10^{-5}$.  Inset: Dependence of conductivity on regularization constant $\lambda$ for fixed mass $m=0.05$ eV.  Both plots are for long-range interactions with $\epsilon=1$ (suspended graphene). }
        \label{fig:mass_and_lambda}
\end{figure}

With this test calculation we should reproduce analytical result for $\sigma(\omega)$ on honeycomb lattice and the figure  \ref{fig:cond_free} indeed demonstrates all essential properties of free spectral function $\sigma(\omega)$: transport peak at zero frequency, then plateau around $\sigma_0$, then van Hove singularity and final decay at large frequencies.  We conclude that we should take resolution function with maximum around $\omega_c \approx \kappa$ to reproduce right value of  optical conductivity. This $\omega_c$ is large enough to get rid of the gap induced by finite mass (we do not see any dependence on bare mass, see below) and small enough for results not to be influenced by van Hove singularity.

In real Monte-Carlo calculations statistical errors  are sufficiently larger and we need some regularization $\lambda \neq 1$. Figure  \ref{fig:conductivity_omega_int} shows optical conductivity $\bar  \sigma(\omega)$ and corresponding resolution functions for suspended graphene ($\epsilon=1$, long-range interaction). Since we are forced to introduce regularization in this case, the width of  $\delta(\omega_0, \omega)$ functions for large $\omega_0$ increases and we can not catch van Hove singularity in our data. Nevertheless, width of $\delta(\omega_0, \omega)$ function with maximum at $\omega_c \approx \kappa$ is rather stable and we can still use it to estimate the value of $\sigma(\omega)$ at plateau which corresponds to optical conductivity of Dirac quasiparticles. This resolution function with maximum at $\omega_c = 1.06 \kappa$ is used in all subsequent calculations.

\begin{figure}[t!]
        \centering
        \includegraphics[scale=0.281, angle=0]{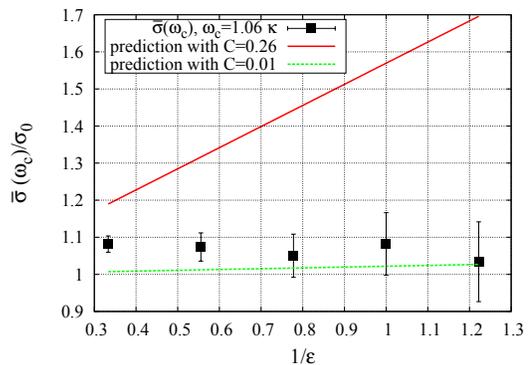}
        \caption{Dependence of $\bar \sigma (\omega_c)|_{\omega_c=1.06 \kappa}$ on interaction strength in the case of long-range interactions. All calculations were performed for $\lambda=1-4\times 10^{-5}$ and $m=0.05$ eV.  Two variants of analytical predictions for graphene conductivity renormalization (see eq. \ref{sigma_renorm}) are plotted for reference.
        }
        \label{fig:conductivity_alpha}
\end{figure}

Now we should check independence of results on bare mass in the Hamiltonian (\ref{tbHam}) and regularization  $\lambda$. Figure \ref{fig:mass_and_lambda} demonstrates results of this check in the case of full calculation in interacting model.  Main plot shows that we don't have any definite dependence on bare mass. Thus we simply use the lowest possible $m=0.05$ eV in all calculations.  Effect of regularization is shown in the inset. Again the dependence on $\lambda$ is flat unless the instability of analytical continuation causes sharp growth of errors at $\lambda \rightarrow 1$. According to results of this test we choose regularization constant $\lambda=1-4\times10^{-5}$.

\begin{figure}[t!]
        \centering
        \includegraphics[scale=0.27, angle=0]{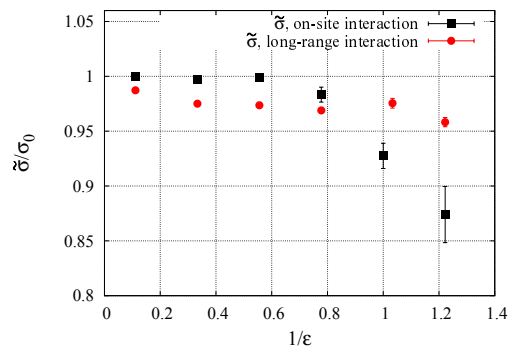}
        \caption{Left scale: dependence of  conductivity $\tilde \sigma$ (calculated from the middle point of current-current correlator, see eq. \ref{corr_midpoint})  on the interaction strength both for long-range and short-range interactions. Phase transition into antiferromagnetic state is situated at $\varepsilon=0.6...0.7$ in the case of long-range interactions (see \cite{Ulybyshev:2013swa}) and at $\varepsilon = 0.91$ (see \cite{Assaad:2013}) in the case of pure on-site interaction.
        }
        \label{fig:conductivity_midpoint}
\end{figure}

Using this set up we plot  the dependence of conductivity on interaction strength in the case of long-range interaction (fig. \ref{fig:conductivity_alpha}). For reference we plot at the same figure also two variants of analytical predictions according to eq. \ref{sigma_renorm} for two different scenarios: large renormalization with $C=0.26$ and small renormalization $C=0.01$. Our data is consistent only with small renormalization scenario while large renormalization of graphene conductivity is completely ruled out by our calculations.

In order to estimate conductivity in the case of short-range interactions, we used alternative approach.  Unfortunately, statistical errors in current-current correlator are too large in this case and we are unable to extract $\sigma(\omega)$ directly from Green-Kubo relations. We used estimate of optical conductivity from the middle point of correlator (taken in the massless limit):
\beq
\tilde \sigma = \frac{\beta^2} {\pi} G(\beta /2)|_{m \rightarrow 0}.
\label{corr_midpoint}
\eeq
 The method was proposed in \cite{Buividovich:2012nx}. Physically it corresponds to conductivity smeared over some region of frequencies near $\omega=0$. Width of the region is approximately equal to $2T$:
 \beq
\tilde \sigma = \frac{\beta^2} {\pi} \int_0^\infty \frac{d\omega}{2 \pi}\frac{2 \omega}{\sinh\left({\frac{1}{2}\beta \omega}\right)} \sigma (\omega).
\label{midpoint_explain}
\eeq
 In the case of free fermions this approach gives us the value  $\tilde \sigma  \approx 0.28$ which is also rather close to $\sigma_0$. Results for conductivity calculation from the middle point of current-current correlator ($\tilde \sigma$) are shown in the figure \ref{fig:conductivity_midpoint} in the case of short- and long-range interactions. Again, this method doesn't show any sufficient renormalization of optical conductivity in the case of long-range interaction. In the case of short-range interactions we see decrease of conductivity for large interaction strength.
 
Stronger renormalization in the case of pure short range interaction is connected to the fact that antiferromagnetic (AFM)  phase transition is closer to the $\varepsilon=1$ point in the case of short-range interaction. While it takes place at $\varepsilon=0.6...0.7$ in the model with long-range Coulomb tail \cite{Ulybyshev:2013swa}, pure on-site interaction  causes phase transition at $V_{00}=3.78 \kappa$ \cite{Assaad:2013} which corresponds to $\varepsilon=0.91$ in our notation (see the description of the interactions after eq. (\ref{tbHam})). One can conclude that decrease of $\tilde \sigma$ in the case of short-range interactions is connected to proximity to the AFM phase transition. Thus, our calculations are accurate enough to agree with analytical results \cite{Mastropietro}; beyond the phase transition, perturbative arguments of that work are no more valid. 

\section{Conclusions}

To conclude, straightforward Quantum Monte Carlo calculations of optical conductivity for interacting electrons on honeycomb lattice show a very weak dependence on the coupling constant. Numerical data completely rules out the ``strong renormalization'' scenario.  It can be consistent with both an accidental ``almost cancellation'' of self-energy and vertex corrections leading to a small coefficient $C$ in Eq.(\ref{sigma_renorm}) and an exact calculation due to Ward identity, similar to the case of Hubbard model \cite{Mastropietro}. It is not clear yet, however, why such exact cancellation should take place for long-range interactions. We hope that computational results presented here will stimulate further analytical studies of this very complicated and controversial issue.

\section*{Acknowledgements}
The work of MU was supported by the DFG Grant BU 2626/2-1 and by
Grant RFBR-14-02-01261-a. MIK acknowledges financial support from NWO
via Spinoza Prize. The work of DLB was supported by RFBR grant
16-32-00362-mol-a. The work of VVB, which consisted of the calculation of the
current-current correlator, was supported by grant from the
Russian Science Foundation (project number 16-12-10059).  The authors
are grateful to FAIR-ITEP supercomputer center and to supercomputer
center of Moscow State University.

\end{document}